\documentclass[final,5p,twocolumn]{elsarticle}
\usepackage{amsthm}
\usepackage{natbib}
\usepackage{xcolor}
\usepackage{amssymb}
\usepackage{graphicx}
\usepackage{subfigure}
\theoremstyle{definition}
\usepackage[page]{appendix}

\newtheorem{definition}{Definition}

\begin{document}
\begin{frontmatter}
\title{Mining Area Skyline Objects from Map-based Big Data using Apache Spark Framework}                   

\author[label1]{Chen Li}
\affiliation[label1]{
organization={Graduate School of Informatics, Nagoya University},
city={Chikusa, Nagoya},
postcode={464-8602}, 
country={Japan}}
\ead{li.chen.z2@a.mail.nagoya-u.ac.jp}

\author{Ye Zhu \corref{cor1}\fnref{label2}}
\affiliation[label2]{
organization={Deakin University},
addressline={Centre for Cyber Resilience and Trust}, 
city={Burwood},
postcode={3125}, 
country={Australia}}
\ead{ye.zhu@ieee.org}

\author[label2]{Yang Cao}
\ead{charles.cao@ieee.org}

\author[label3]{Jinli Zhang}
\affiliation[label3]{
organization={Beijing University of Technology},
city={Beijing},
postcode={100124}, 
country={China}}
\ead{lz73798@gmail.com}

\author[label4]{Annisa Annisa}
\affiliation[label4]{
organization={Department of Computer Science, IPB University},
country={Indonesia}}
\ead{annisa@apps.ipb.ac.id}

\author[label5]{Debo Cheng}
\affiliation[label5]{
organization={UniSA STEM, University of South Australia},
country={Australia}}
\ead{chedy055@mymail.unisa.edu.au}


\author[label6]{Yasuhiko Morimoto}
\affiliation[label6]{
organization={Graduate School of Engineering, Hiroshima University},
city={Higashi-hiroshima},
postcode={10587}, 
country={Japan}}
\ead{morimo@hiroshima-u.ac.jp}

\cortext[cor1]{Corresponding authors}

\begin{abstract}
The computation of the skyline provides a mechanism for utilizing multiple location-based criteria to identify optimal data points. However, the efficiency of these computations diminishes and becomes more challenging as the input data expands. This study presents a novel algorithm aimed at mitigating this challenge by harnessing the capabilities of Apache Spark, a distributed processing platform, for conducting area skyline computations. The proposed algorithm enhances processing speed and scalability. In particular, our algorithm encompasses three key phases: the computation of distances between data points, the generation of distance tuples, and the execution of the skyline operators. Notably, the second phase employs a local partial skyline extraction technique to minimize the volume of data transmitted from each executor (a parallel processing procedure) to the driver (a central processing procedure). Afterwards, the driver processes the received data to determine the final skyline and creates filters to exclude irrelevant points. Extensive experimentation on eight datasets reveals that our algorithm significantly reduces both data size and computation time required for area skyline computation.
\end{abstract}

\end{frontmatter}

\section{Introduction}
\label{sec:intro}
Identifying optimal locations on a map holds significant importance across diverse applications such as trip planning \cite{kar2023location}, real estate \cite{li2005gis}, and business analysis \cite{smirnov2020context}. The rise in popularity of map-based mobile applications in recent years is attributed to their capacity to assist users in navigating maps and effortlessly identifying points of interest \cite{bartling2022adapting}. Nevertheless, the wealth of information displayed on maps can pose a challenge for users, leading to difficulty in selecting the most suitable locations and potentially causing them to feel overwhelmed by the array of options available.

An area skyline query, as defined by Zaman \cite{zaman2016area}, serves as a spatial database query crucial for location-based decision-making. It retrieves a subset of objects within a specified area, distinguished by their non-domination in comparison to other locations. In essence, this query identifies regions within a given area, each possessing at least one attribute that surpasses that of any other region \cite{djatna2020implementation}. To execute this query, the database initially divides the entire space into a grid of cells, with the area of interest specified by the query, potentially spanning multiple cells. Subsequently, within the defined area, the database retrieves all objects, conducting thorough comparisons among them. The outcome is the area skyline, consisting of non-dominated objects. The challenge inherent in area skyline queries lies in their high computational complexity, particularly evident as the number of objects and the size of the query area increase. This growth results in a proportional increase in query execution time. To mitigate this computational burden, optimization algorithms such as MapReduce can be effectively employed for parallel processing, significantly improving the efficiency of area skyline query performance.

MapReduce is a widely-utilized programming model in distributed computing, providing a scalable and resilient framework for processing extensive datasets \cite{bashabsheh2022big, park2013parallel, eldawy2015spatialhadoop}. Its effectiveness in managing intricate operations on large datasets is particularly noteworthy. An illustration of such complexity is the area skyline query, a task seamlessly achievable through MapReduce \cite{li2017mapreduce}. In a MapReduce-based implementation of the area skyline query, data undergoes initial partitioning and distribution across multiple nodes within a cluster. Subsequently, the Map function is applied to each partition, extracting relevant object attributes and discerning the local skyline within that specific partition. The local skylines are then amalgamated using the Reduce function to derive the global skyline for the entire dataset \cite{li2018mapreduce}.

The utilization of MapReduce for area skyline queries offers various advantages, notably its capability to manage extensive datasets through parallel processing, consequently reducing computation time. Moreover, a significant benefit of employing the MapReduce framework for area skyline computation lies in its inherent fault tolerance.  In a distributed computing environment, it is not uncommon for individual nodes in the cluster to encounter failures or disruptions. However, MapReduce is designed to handle such situations seamlessly \cite{bessani2010making}. 

Apache Spark \cite{pallamala2022investigative,salloum2016big} is a distributed computing framework that has gained significant popularity in recent years for large-scale data processing. Compared to MapReduce, Apache Spark provides improved performance and greater flexibility \cite{guha2023ngs,shi2015clash}. Some of the key advantages of using Apache Spark over MapReduce are listed below \cite{shi2015clash,garcia2017comparison}:

\begin{itemize}
\item{\bf Speed:} Apache Spark is faster than MapReduce primarily because of its in-memory processing capability. MapReduce stores data on disk after each operation, resulting in delays during data processing. In contrast, Apache Spark keeps data in memory, allowing it to perform computations significantly faster as data can be quickly accessed and manipulated without having to read and write from disk.
\item{\bf Ease of use:} Apache Spark provides a programming model that is simpler than MapReduce, enabling developers to write and maintain code with ease. Additionally, Spark also offers APIs for different programming languages, like Java, Python, and Scala, making it even more convenient for developers to develop their applications.
\item{\bf Real-time processing:} Apache Spark is designed to support real-time data processing, which makes it an excellent choice for applications that require low-latency data processing, such as online recommendations or fraud detection.
\item{\bf Advanced analytic:} Apache Spark provides a wide range of libraries for advanced analytic, including machine learning, graph processing, and stream processing. These libraries enable developers to perform complex data analysis tasks efficiently.
\item{\bf Memory management:} Apache Spark owns better memory management capabilities than MapReduce, allowing it to handle larger datasets without running out of memory.
\item{\bf Fault tolerance:} Similar to MapReduce, Apache Spark also offers fault tolerance capabilities, but with faster recovery times than MapReduce. In the event of a node failure, Spark can recover data from memory or disk, and continue processing without restarting the whole job.
\end{itemize}

The current MapReduce-based approaches for area skyline algorithms involve computing the distance between each grid and the facility \cite{li2017mapreduce,li2018mapreduce}. However, these methods lack support for distance tuples and skyline calculations within distributed frameworks. In this investigation, we present an enhanced algorithm built on the Apache Spark framework to address area skyline queries. The primary contributions of our study are outlined as follows:
\begin{itemize}
\item {\bf Development of an Apache Spark-based distributed algorithm:} Apache Spark, a distributed framework for big data, is utilized to create an algorithm that offers improved performance and flexibility compared to MapReduce.
\item {\bf Introduction of a distributed Area skyline calculation:} This study puts forth a distributed algorithm designed for computing distance tuples and area skylines within the Apache Spark framework. In contrast to prior MapReduce-based approaches for area skyline computation, the proposed algorithm seamlessly integrates skyline selection and filtering.
\item {\bf Demonstration of superior performance compared to State-of-the-Art (SOTA) baselines:} Experimental results illustrate that our proposed distributed algorithm exhibits enhanced performance and calculation speed in comparison to SOTA models.
\end{itemize}

The rest of our paper is organized as follows: Section 2 reviews the related work of skyline query, as well as skyline Computation MapReduce. Section 3 introduces examples of the area skyline and the distributed algorithm for area skyline computation. Section 4 describes the proposed algorithm of area skyline computation with the Apache Spark framework. Section 5 reports and analyzes an empirical evaluation of the proposed distributed area skyline calculation. We make the conclusion and future work in Section 6.

\section{Related Work}
\label{sec:related}

In this section, we will discuss some of the important literature related to our proposed method, including skyline query, skyline computation with the distributed framework and area skyline computation with MapReduce.  
\subsection{Skyline Query}
\label{sec:skyline}
\cite{borzsony2001skyline} first introduced skyline computation with three different algorithms: the block nested loop (BNL), the Divide-and-Conquer (D$\&$C), and the B-tree-based algorithms). The BNL algorithm computes the skyline, i.e., the non-dominated points, in a dataset. It compares each point with all others to check for dominance. The steps include retrieving the dataset, initializing an empty result set, processing each point, comparing it with others, performing a dominance check, adding non-dominated points to the result set, and returning the final skyline. The BNL algorithm is simple but has a time complexity of $O(n^2)$, making it suitable for small to medium-sized datasets, while more efficient algorithms are recommended for larger datasets. The D$\&$C algorithm is a technique for solving problems by breaking them into smaller sub-problems, solving each sub-problem independently, and then combining the results. In the context of skyline computation, the D$\&$C algorithm recursively partitions the dataset into smaller subsets, computes the skyline for each subset, and then merges the skylines to obtain the final result. D$\&$C can reduce the overall computational complexity compared to the BNL algorithm. The algorithm repeatedly divides the dataset, conquers each partition, and combines the skylines until the entire dataset is processed. It achieves improved efficiency for larger datasets by exploiting the divide-and-conquer strategy. The B-tree-based algorithm is a technique for efficient indexing and searching in large datasets. It utilizes a balanced tree structure called a B-tree to store and organize the data. The B-tree allows for efficient insertion, deletion, and retrieval operations by maintaining a balanced height and minimizing disk accesses. In the context of skyline computation, the B-tree-based algorithm constructs an index structure on the dataset and performs efficient pruning and traversal operations to identify the skyline points. This approach reduces the search space and optimizes the skyline computation process, making it suitable for handling large datasets with improved performance compared to other algorithms.

In the same year, \cite{tan2001efficient} proposed the Bitmap and Index (BI) algorithm. The BI algorithm combines the advantages of bitmap indexing and traditional index structures. The algorithm constructs bitmap indexes for each attribute dimension to represent dominant relationships between points. It also creates auxiliary index structures, such as B-trees, to facilitate efficient pruning and traversal operations. By leveraging the bitmaps and indexes, the algorithm quickly identifies skyline points by performing bitwise operations and utilizing index lookups. This approach significantly reduces the computational complexity and improves the overall performance of skyline query processing, making it suitable for large and high-dimensional datasets.

To reduce the number of comparisons in BNL, \cite{chomicki2003skyline} then proposed the Sort-Filter-Skyline (SFS) algorithm which was applied to sort steps before skyline computation. The SFS algorithm follows a three-step process: sorting, filtering, and skyline computation. First, the algorithm sorts the dataset based on attribute values in non-decreasing order. Then, it filters out dominated points by iteratively comparing points with previously encountered skyline points. Finally, it computes the skyline by considering the remaining non-dominated points. The SFS algorithm leverages the sorted order of the dataset to reduce the number of necessary dominance checks, resulting in improved efficiency for skyline queries. By filtering out dominated points early in the process, it minimizes unnecessary comparisons, significantly reducing computational overhead. The SFS algorithm is particularly effective for large datasets with high-dimensional attribute spaces, where it can achieve substantial performance gains compared to other skyline query processing techniques.

Using R-tree based indexing, in 2008, \cite{xia2008skylining} proposed the fastest skyline query computation, Branch and Bound Skyline (BBS) algorithm. The BBS algorithm utilizes a branch and bound strategy to minimize the number of unnecessary point comparisons. The algorithm recursively divides the dataset into branches and bounds the search space by pruning dominated branches. It explores the branches based on dominance relationships and incrementally builds the skyline set. By intelligently pruning dominated branches, the BBS algorithm significantly reduces the computational complexity of the skyline query. It is particularly effective for high-dimensional datasets, where it outperforms naive skyline computation methods by efficiently navigating the search space to identify the skyline points.

\subsection{Skyline Processing in Distributed Framework}
\label{sec:skyline_hadoop}

One of the important issues in skyline computation is how to speed up skyline computation. Some researchers proposed parallel-based computation using MapReduce framework to deal with the issue.
In \cite{zhang2011adapting}, MapReduce was first introduced for skyline computation. They introduced MR-BNL and MR-SFS in the study. In general, these two algorithms consist of three steps: calculating center
value, computing local skyline, and calculating global skyline.
To calculate the center value, in the Map process, a tuple of the form $<key, value>$ will be read from HDFS. Furthermore, in the Reduce stage, the center value will be calculated using a list of values for each dimension that is already sorted, and then written to HDFS. The local skyline calculation is performed by the executor using the center value obtained in the previous step. In the next process, the HDFS tuple will be read and divided into two subspaces based on the center value, then each ID and tuple value from each subspace will become a key and value in the MAP process.
The skyline calculation for each executor is calculated based on the set of tuples in the same subspace in the Reduce process. The skyline that is obtained from each executor is called a local skyline, because it does not come from all the existing data.
To get the global skyline, the next Map and Reduce process will read the local skyline from HDFS and include all tuples in the calculation.
The MR-BNL and MR-SFS algorithms above have some drawbacks. In the Reduce process to calculate the local skyline, parallel calculations are only executed for 2 $*$ dimension tasks. If then there is an increase in the number of nodes, the performance of the algorithm will decrease and be difficult to improve. In addition, the burden of calculating the global skyline in Reduce process is also very large because the data is only concentrated on one executor.

To overcome the limitations in MR-BNL and MR-SFS, an algorithm based on MapReduce is proposed, called SKY-MR\cite{park2013parallel}. MapReduce does parallel filtering of the global skylines on SKY-MR. The three stages contained in SKY-MR include building the sky quadtree, calculating the local skyline, and calculating the global skyline.
The removal of tuples that do not belong to any node from the sky quadtree is performed on the Map process. The tuple of a node consists of the ID as the key and the tuple as the value. Skyline calculations on data belonging to the same node occur in the Reduce Process. The obtained local skyline becomes the basis for creating the virtual maximum points and skyfilters.
The virtual tuple whose dimension is the maximum value of the local skyline is called the virtual maximum point.
Minimum values for each dimension in local skyline points are filtered using Skyfilter.
before global skyline calculations, Sky filters and virtual maximum points are sent to the local storage of the driver and each executor from HDFS.
local skyline points which are the skyfilter filter tuples used in global skyline calculations in the Map process.
then, the skyline calculation for points belonging to the same node is performed in the Reduce Process.
The MapReduce framework is also used in MR-Angle by splitting the data for the calculation of the skyline by the angle to the origin
to calculate the skyline query, \cite{chen2012mapreduce}.

Apache Spark framework, which is specifically designed for distributed processing, in recent years has attracted many researchers to use it for executing skyline queries over big and distributed data \cite{papanikolaou2020distributed,grasmann2022integration,lapatta2022ecotourism}. Based on divide-and-conquer approach in SparkALS \cite{papanikolaou2020distributed}, the data is divided into smaller subsets, and from each subset, the skyline is calculated in parallel. To get the final result, all Skylines that have been obtained in the previous stage are then merged. Apache Spark's resilient distributed datasets abstraction on SparkALS greatly supports distributed processing of the skyline computation. The Spark SQL engine is extended by SparkSQL \cite{grasmann2022integration} so that it can be used for skyline query calculations by defining new user defined functions (UDFs). Spark's DataFrame API is then used to implement UDF so that skyline query processing can run in an efficient and scalable distributed environment. A skyline query approach, Senti-Spark \cite{lapatta2022ecotourism} used the Apache Spark cluster computing framework to build recommendations based on the Skyline Sort Filter algorithm. Their algorithm can ensure that the best ecotourism recommendations with positive sentiments are given to tourists, proven by the results of the experiments. 

\subsection{Area Skyline Computation with Hadoop MapReduce}

All the algorithms described previously are algorithms for performing computations to obtain skyline objects. Area skyline is the problem of finding skyline objects in two dimensions. The algorithm used to find the area skyline using MapReduce was proposed in \cite{li2018mapreduce,li2017mapreduce} research.
The algorithm effectively incorporated the Euclidean distance transformation algorithm into the MapReduce framework. In each row of the map, the Map process computes the shortest distance between the grid and the facility. Subsequently, the data undergoes randomization in the Shuffle process before entering the Reduce process. In the Reduce process, the distance between each grid and the facility is determined by identifying the grid with the closest distance to the facility, considering the values of each grid.
The described algorithm primarily emphasizes skyline area calculation, but encounters challenges in improving execution time during the Reduce process, especially when computing distance and local skylines. Consequently, the overall computation process experiences a significant increase in execution time. Furthermore, the execution time for local skyline computation within the Reduce process exhibits notable escalation. Specifically, as the number of facilities and grids increases, the execution time of the Reduce process for local skyline computation experiences extensive growth. In this study, we enhance the distributed algorithm for area skyline computation based on the Apache Spark framework, addressing the issues outlined in the following section.

\section{Preliminary}
\label{sec:preliminary}
This section introduces the concept of area skyline and presents a distributed (i.e., MapReduce-based) algorithm for computing the area skyline.

\subsection{Area Skyline Computation}
The term “area skyline" pertains to a collection of points within a specific region or area of interest that remains unconquered by any other point. Essentially, it offers a localized perspective of the skyline within a defined boundary. To formalize, let $A_{area}$ represent a square area on a map, and let $F = \{F_1, F_2, \cdots\}$ denote a set of facilities, which could be stations, warehouses, or landfills. The user's objective is to identify a region on the map that is proximate to desirable facilities while being distanced from undesirable ones. In this context, the definitions for area skyline and area dominance are articulated as follows:

\begin{definition} 
\it{In the division of a square area $A_{area}$ into $k \times k$ grids, denoted as $A_{area} = \{a_1, a_2,\cdots,a_{k \times k}\}$, a grid $a_i$ qualifies as an area skyline object if and only if there exists no other grid $a_j~(i \neq j)$ on the map for which the distance to the closest desirable facilities is less than that of $a_i$, and simultaneously, the distance to the closest undesirable facilities is greater than that of $a_i$.}
\end{definition} 

\begin{definition} 
\it{In the scenario where a grid $a_i$ is identified as an area skyline object, it implies that $a_i$ either dominates $a_j~(i \neq j)$ or is dominated by $a_i$.}
\end{definition} 

\begin{figure}[ht]
\centering
\includegraphics[width=0.8\hsize]{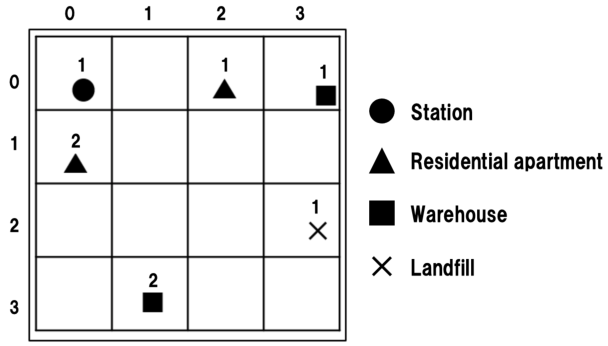}
\caption{Example of area skyline in a map.}
\label{fig:area_skyline_example}
\end{figure}
Figure \ref{fig:area_skyline_example} displays an example of an area skyline. Suppose that a businessman considers the construction of a new commercial facility on the map. The map is divided into a total of 16 grids, and four types of facilities, consisting of a train station, a residential apartment building, a warehouse, and a refuse disposal facility, are scattered on the grids. Among these four types of facilities, the station, residential apartments, and the company's warehouse are favorable facilities, while the refuse disposal facility is unfavorable. Grid 00 has the minimum distance to the station (distance to station 1) of 0, the minimum distance to the apartment house (distance to apartment house 2) of 1, the minimum distance to the company warehouse (distance to company warehouse 1) of 3, and the minimum distance to the waste disposal site (distance to waste disposal site 1) of $\sqrt{13}$. Similarly, the minimum distances to other grids and facilities are shown in Table \ref{tab:area_skyline_example}. G02 is better than G01 in the minimum distance to the residential apartments. However, G01 is better than G02 for the distance to the landfill. In addition, G23 is worse than G20 in the minimum distance to all facilities. On this map, the set of grids that are not worse than all other grids is $\{G00, G01, G02, G03, G10, G20, G30, G31\}$, and this set is extracted as the area skyline. Therefore, we can conclude that only the grids included in the area skyline should be considered when constructing a new facility.

Figure \ref{fig:area_skyline_example} illustrates an example of the area skyline concept. Let's consider a businessman planning to construct a new commercial facility on a map divided into 16 grids. The grids contain four types of facilities: train stations, residential apartments, warehouses, and a refuse disposal facility. Among these facility types, the stations, residential apartments, and warehouses are considered favorable, while the refuse disposal facility is unfavorable. Each grid has various distances to the different facilities. For example, Grid 00 has the minimum distance to the station (distance to station 1) of 0, the minimum distance to the apartment house (distance to apartment house 2) of 1, the minimum distance to the company warehouse (distance to company warehouse 1) of 3, and the minimum distance to the waste disposal site (distance to waste disposal site 1) of $\sqrt{13}$. Similarly, the minimum distances to other grids and facilities are shown in Table \ref{tab:area_skyline_example}. From the table, G02 is better than G01 in terms of the minimum distance to residential apartments, but G01 is superior when considering the distance to the landfill. Additionally, G23 is worse than G20 in terms of the minimum distance to all facilities. In this scenario, the area skyline is the set of grids that are not worse than any other grid. In this example, the area skyline consists of $\{G00, G01, G02, G03, G10, G20, G30, G31\}$. Therefore, when constructing a new facility, only the grids included in the area skyline should be considered. The area skyline provides valuable insight into identifying optimal locations for constructing facilities, taking into account the distances to various types of facilities. By focusing on the area skyline, the businessman can make informed choices to optimize the placement of new constructions.

\begin{table*}[ht]
\setlength\tabcolsep{6pt}
\centering
\caption{An example of the computation of area skyline attributes.}
\begin{tabular}{ccccc}\hline
{\bf Grid} & {\bf Station} ($\downarrow$) & {\bf Apartment} ($\downarrow$) & {\bf Warehouse} ($\downarrow$) & {\bf Landfill} ($\uparrow$)\\\hline \hline 
$G_{00}$ & 0 & 1 & 3 & $\sqrt{13}$ \\
$G_{01}$ & 1 & 1 & 2 & $2\sqrt{2}$ \\
$G_{02}$ & 2 & 0 & 1 & $\sqrt{5}$ \\
$G_{03}$ & 3 & 1 & 0 & 2 \\
$G_{10}$ & 1 & 0 & $\sqrt{5}$ & $\sqrt{10}$ \\
$G_{11}$ & $\sqrt{2}$ & 1 & 2 & $\sqrt{5}$ \\
$G_{12}$ & $\sqrt{5}$ & 1 & $\sqrt{2}$ & $\sqrt{2}$ \\
$G_{13}$ & $\sqrt{10}$ & $\sqrt{2}$ & 1 & 1 \\
$G_{20}$ & 2 & 1 & $\sqrt{2}$ & 3 \\
$G_{21}$ & $\sqrt{5}$ & $\sqrt{2}$ & 1 & 2 \\
$G_{22}$ & $2\sqrt{2}$ & 2 & $\sqrt{2}$ & 1 \\
$G_{23}$ & $\sqrt{13}$ & 5 & 2 & 0 \\
$G_{30}$ & 3 & 2 & 1 & $\sqrt{10}$ \\
$G_{31}$ & $\sqrt{10}$ & $\sqrt{5}$ & 0 & $\sqrt{5}$ \\
$G_{32}$ & $\sqrt{13}$ & $2\sqrt{2}$ & 1 & $\sqrt{2}$ \\
$G_{33}$ & $3\sqrt{2}$ & $\sqrt{10}$ & 2 & 1 \\\hline
\end{tabular}
\label{tab:area_skyline_example}
\end{table*}

\subsection{Distributed MapReduce Algorithm for Area Skyline Computation}
The MapReduce-based area skyline query is a computational approach that leverages the MapReduce framework to efficiently compute the area skyline in large-scale datasets. The area skyline refers to the set of non-dominated points within a specified region of interest. Generally, the MapReduce-based area skyline algorithm can be divided into the following two parts:
\begin{figure}[t]
\centering
\subfigure[Map function for area skyline compuation.]{
\includegraphics[width=0.33\textwidth]{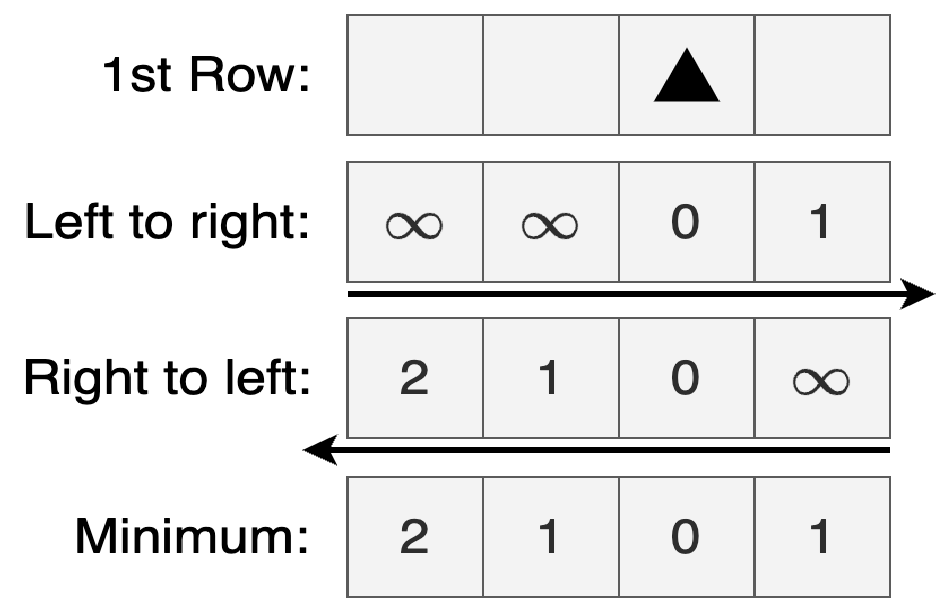}
\label{fig:map}
}
\subfigure[Remove $p_j$ in the Reduce function.]{
\includegraphics[width=0.33\textwidth]{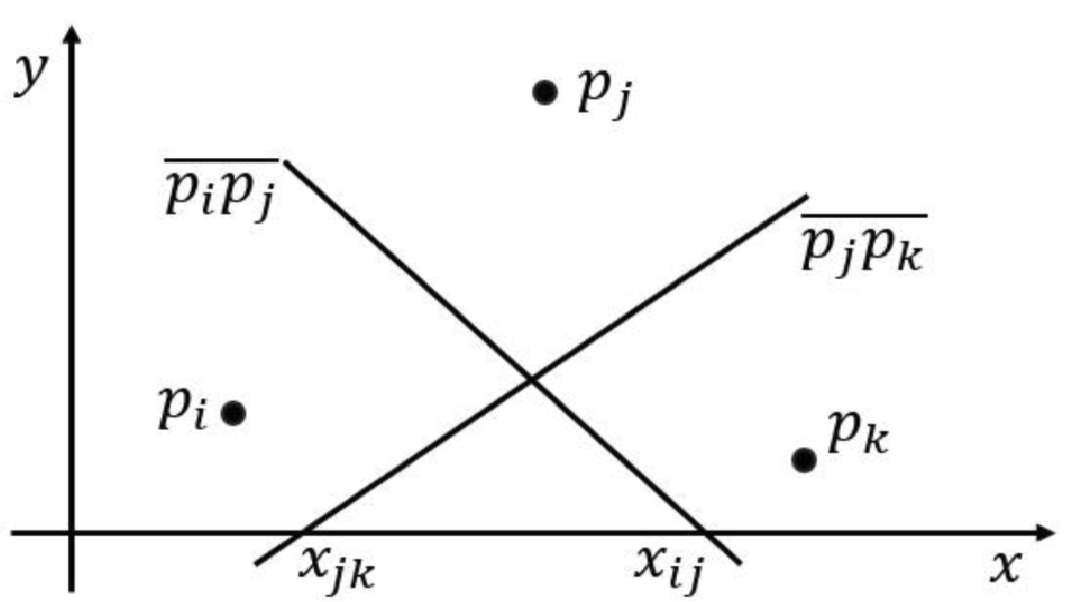}
\label{fig:delete}
}
\subfigure[Maintain $p_j$ in the Reduce function.]{
\includegraphics[width=0.33\textwidth]{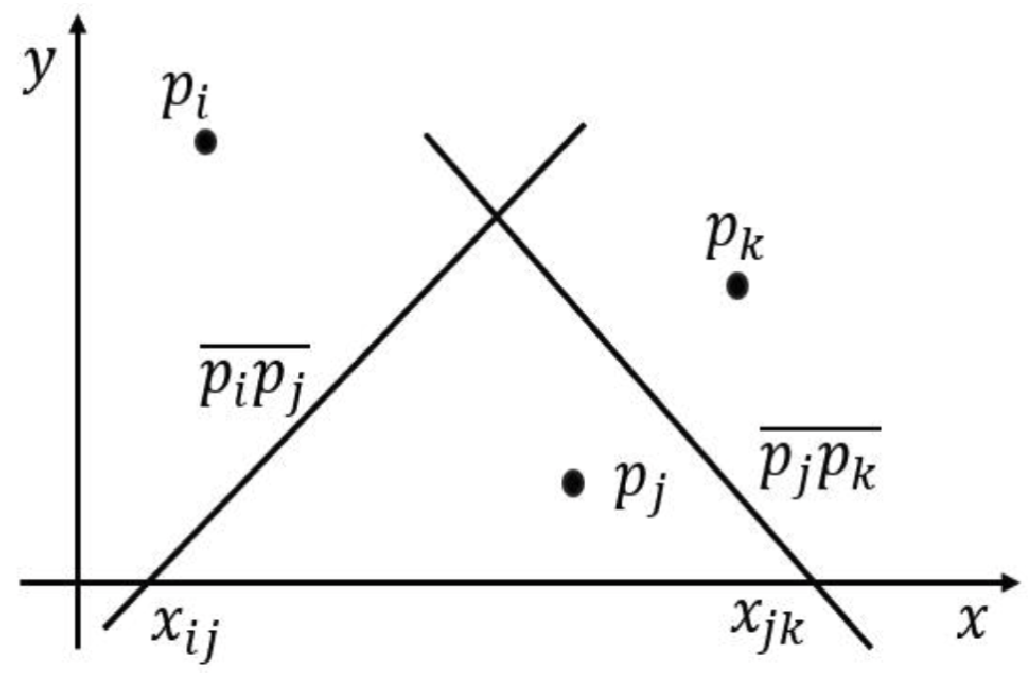}
\label{fig:maintain}
}
\caption{An example of a distributed algorithm for area skyline computation.}
\end{figure}

\begin{itemize}
\item \textbf{Map function} computes the distance to the nearest facilities of each type within the same row. In this function, the map is initially divided into $k$ rows. For each row, the Map function scans the grids from both ends to determine the distance between the grids and the facilities. At the outset, all grid values are assumed to be infinite. Upon encountering facilities, the grid values are set to 0. Subsequently, the value of the next grid is updated incrementally from the preceding grid until the next facility is reached. Fig. \ref{fig:map} (upper panel) provides an example where shaded grids represent facilities of type F1+. The values of these shaded grids are set to 0. Progressing from left to right, the values of the 5th, 6th, and 8th grids are successively updated to 1, 2, and 1. A similar computation is performed from right to left. After computing values from both directions, the output of the Map function is determined by selecting the minimum value for each grid, as illustrated in Fig. \ref{fig:map} (lower panel).

\item In the \textbf{Reduce function}, the computation involves determining the distance of each grid to the nearest facilities within the same column. This process begins by sorting and shuffling the key-value pairs generated by the Map process based on the column. The Reduce function then iterates through the key-value pairs in the ith column from bottom to top, storing the values in a stack. It's crucial to emphasize that each column has its dedicated stack. To facilitate this analysis, the grids within the same column are represented in a two-dimensional coordinate system. Here, the x-axis corresponds to the column IDs of the grids (enumerated from bottom to top), and the y-axis reflects the outcomes from step 1.2. Each point is denoted as $p_1, p_2, \cdots, p_n$. The subsequent step involves bisecting adjacent points $p_i=(x_i, y_i)$ and $p_j =(x_j, y_j)$ where $(1 \le i < j \le n)$. The intersection of the vertical bisector $\hat{p_ip_j}$ and the x-axis is computed as $x_{ij}$. The calculation of $x_{ij}$ is given by
$x_{ij}=\frac{(y_j^2-y_i^2)+(x_j^2-x_i^2)}{2(x_j-x_i)}$. Consider three consecutive points, $p_i$, $p_j$, and $p_k$, with $(i < j < k)$. The decision to remove $p_j$ from the stack hinges on whether $x_{ij} > x_{jk}$. If this condition holds, $p_j$ is deleted; otherwise, it is retained in the stack. The two scenarios, one involving the deletion of $p_j$ and the other maintaining it, are depicted in Figures \ref{fig:delete} and \ref{fig:maintain}. Subsequently, the distance of the removed points from the stack is computed. To achieve this, the adjacent left points are bisected to identify proximate intervals \cite{man2013accelerating} on the x-axis. The interval $[a, b]$ $(a, b \in Z$ and $a \le b)$ of the left point $p_i$ after the process delineates the dominated areas associated with point $p_i$.    

\end{itemize}

By distributing the computation across multiple nodes in a parallel and distributed manner, the MapReduce-based approach enhances scalability and efficiency in skyline query processing. It allows for the effective utilization of computational resources and enables skyline computation on large-scale datasets. The MapReduce-based area skyline query is particularly beneficial when dealing with big data scenarios, where traditional skyline computation algorithms may encounter scalability issues. It provides an optimized and distributed solution for identifying the non-dominated points within a specific area of interest, facilitating decision-making processes in various domains such as urban planning, facility location, and resource allocation.

\section{Area Skyline Computation with Apache Spark Framework}
\label{sec:model}
\begin{figure*}[t]
\centering
\includegraphics[width=0.77\hsize]{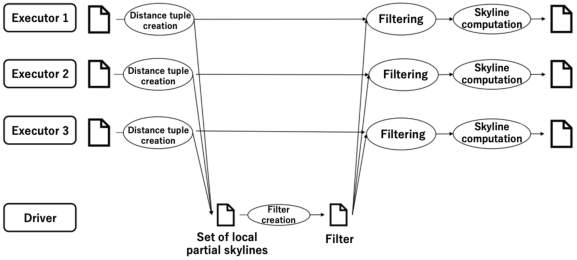}
\caption{Overview architecture of the proposed Apache Spark-based area skyline computation algorithm.}
\label{fig:overview}
\end{figure*}

The architecture of the proposed Apache Spark framework, shown in Figure \ref{fig:overview}, provides an overview of its structure. The algorithm introduces an approach where a subset of the local skylines is extracted during the creation of distance tuples. Additionally, a filter is generated based on these extracted tuples, enabling filtering before the area skyline calculation. The algorithm is divided into three distinct processes:

\begin{itemize}
    \item {\bf Local partial skyline extraction}: this process involves extracting the partial skyline at each executor, specifically computing the skyline for the assigned subset of data.
    \item {\bf Filter creation at the driver}: the algorithm creates a filter using the extracted tuples. This filter serves as an optimization mechanism for subsequent area skyline calculations.
    \item {\bf Filtering in each executor}: each executor employs the created filter to perform selective filtering on the local skylines, eliminating dominated points. This step reduces the computational overhead during the area skyline calculation.
\end{itemize}

\subsection{Local Partial Skyline Extraction}
\label{sec:local}
Within this process, each executor discerns two types of tuples: one with the minimum distance from the origin and another with the minimum value in a specific dimension. These tuples extracted from the executors serve as representations of local skyline points. Consequently, the ensemble of these two types of tuples constitutes a local skyline, representing a subset of the broader local skyline.

\begin{figure}[ht]
\centering
\includegraphics[width=1\hsize]{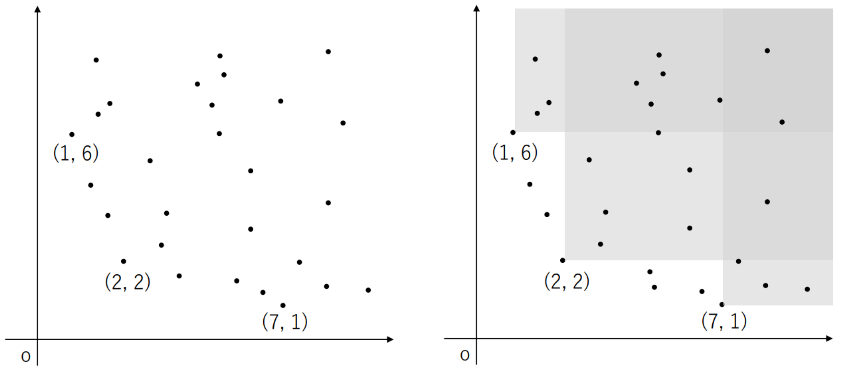}
\caption{An example of local partial skylines and their dominant areas.}
\label{fig:local}
\end{figure}

In a set of tuples, the origin tuple possesses the ability to dominate all other tuples. Consequently, as the distance from the origin decreases, more tuples become susceptible to domination. Additionally, tuples with minimum values in a particular dimension have the potential to dominate other tuples that are not already dominant. For instance, in the left panel of Fig. \ref{fig:local}, a two-dimensional tuple set is displayed along with their corresponding coordinates. The gray region in the right figure represents the area dominated by the respective local partial skyline points. The point closest to the origin, (2, 2), exhibits the largest dominant area. The points (1, 6) and (7, 1), with minimum values in their respective dimensions, dominate the region not already dominated by (2, 2). These three points are referred to as local partial skyline points, which effectively dominate a significant number of tuples in the figure. Typically, a small number of local partial skylines can efficiently filter and eliminate the majority of dominated tuples.

In the proposed algorithm, while creating distance tuples, the process goes beyond merely generating these tuples. It also involves computing and extracting local partial skylines. Upon the formation of a distance tuple, the algorithm scrutinizes both its distance from the origin and the value of each dimension. This examination is conducted to ascertain whether the tuple qualifies for inclusion in the local partial skyline.

\subsection{Filter Creation at Driver}
\label{sec:filter_creation}
In this procedure, the driver node receives local partial skylines from each executor and engages in skyline calculations to generate a filter. When there are $n$ facilities on the map, the tuples become $n$-dimensional, while the local partial skyline points acquired from each executor are $n+1$ dimensional. With a total of $m$ executors, the collective count of local partial skyline points from all executors accumulates to $m \times (n + 1)$. The driver node executes skyline calculations on these $m \times (n + 1)$ tuples.

For instance, in a map with $1000 \times 1000$ grids, featuring 3 facilities and 32 executors, the total number of local partial skyline points obtained from all executors is 128. In constructing the filter, the proposed algorithm's driver node conducts a skyline calculation for these 128 tuples. In contrast, prior approaches necessitated each executor to perform approximately $32$ skyline computations for 1000 tuples during the Reduce process of local skyline computation. Consequently, compared to the Reduce process for local skyline computation, this filtering process imposes substantially lower computational demands and does not serve as a bottleneck for the overall area skyline computation.

\subsection{Filtering in Each Executor}
\label{sec:filtering}
During this process, each executor receives a filter from the driver and utilizes it to filter the tuples before conducting the skyline computation. Initially, the driver broadcasts the filter to all executors. Subsequently, each executor that receives the filter performs dominance checks on each input tuple for the skyline calculation against the tuples in the filter. If a tuple is found to be dominated by any tuple in the filter, it is removed from further consideration. On the other hand, if a tuple is not dominated by any tuples in the filter, it passes through the filter and is retained as input data for the subsequent skyline computation. This filtering step ensures that only non-dominated tuples are considered for the skyline calculation, optimizing the computational resources and reducing the number of comparisons required. By removing dominated tuples early in the process, the algorithm focuses on the most relevant and significant data points, enhancing the efficiency and accuracy of the overall skyline computation.

In summary, the proposed Apache Spark-based algorithm harnesses the power of three essential processes—local partial skyline extraction, filter creation at the driver, and filtering in each executor—to deliver efficient computation of the area skyline. The first process, local partial skyline extraction, plays a critical role in identifying significant skyline points. Each executor is responsible for finding tuples with the minimum distance from the origin and tuples with the minimum value in a specific dimension. These tuples, known as local skyline points, are representative of the local skyline and contribute to the overall area skyline computation. By selectively extracting these skyline points, the algorithm effectively reduces the number of tuples that need to be considered in subsequent calculations, optimizing the computational efficiency. The second process involves filter creation at the driver. The driver node receives the local partial skylines from all executors and combines them to create a filter. This filter, composed of the extracted skyline points, serves as a valuable tool for subsequent computations. By leveraging the filter, the algorithm can significantly reduce the search space during the skyline calculation, as it only needs to evaluate tuples that have the potential to contribute to the final area skyline. This process aids in improving the overall efficiency and effectiveness of the computation. The third process, filtering in each executor, further refines the data by eliminating tuples that are dominated by others. Each executor takes in a filter from the driver and then filters its input tuples according to the dominance relationship. Non-dominated tuples in the filter are processed through the filter, and skyline computation is performed, discarding any dominated tuples. This filtering step helps in reducing the computational burden by focusing on the most relevant and significant tuples, enhancing the efficiency of the overall area skyline computation. By integrating these three processes, the proposed algorithm leverages the benefits of partial skyline extraction, filter creation, and filtering operations to achieve efficient computation of the area skyline. It effectively reduces the search space, eliminates dominated tuples, and focuses on the most relevant skyline points, leading to faster and more accurate results.

\section{Experiments}
\label{sec:exp}
In this section, we conduct comprehensive experiments to thoroughly validate the effectiveness and performance of the proposed Apache Spark framework for area skyline computation. The primary objective of these experiments is to assess the efficiency, scalability, and accuracy of the proposed framework in handling large-scale datasets and computing the area skyline in a timely manner.
\subsection{Experimental Configuration}
\label{sec:conf}
To streamline the setup process and maximize computational efficiency, we leveraged Azure HDInsight \cite{chauhan2014introducing}, provided by Microsoft as a cloud service. This service allowed us to effortlessly create Hadoop and Spark clusters without the need for manual environment configuration. By utilizing Azure HDInsight, we were able to expedite the creation of Spark clusters for our experimental setup. The configuration details of the clusters used in our experiments are outlined in Table \ref{tab:exp_statistics}. To ensure optimal resource allocation and efficient parallel processing, each distributed task is allocated a memory resource of 7GB for each core. This allocation strategy guaranteed sufficient resources for the parallel execution of tasks, facilitating efficient data processing and analysis. Specifically, the Apache Spark application was tailored to employ 24 parallel distributed executors. This configuration enabled high-performance processing and analysis of data, leveraging the parallel computing capabilities of Apache Spark. By harnessing the power of parallelism, we aimed to achieve faster and more efficient computation of the area skyline. By utilizing Azure HDInsight, we simplified the process of creating and managing Spark clusters, allowing us to focus on the experimental evaluation of the proposed Apache Spark framework. The seamless integration with Azure HDInsight optimized resource allocation, enhanced scalability, and maximized computational efficiency, enabling us to conduct comprehensive experiments and validate the effectiveness of the proposed framework.

\begin{table}[ht]
\setlength\tabcolsep{8pt}
\centering
\caption{Configuration of Apache Spark clusters.}
\begin{tabular}{cccc}\hline
Node type & RAM & \# of cores & \# of machines \\\hline \hline 
Master node & 64GB & 8 & 2 \\
Worker node & 64GB & 8 & 3 \\\hline
\end{tabular}
\label{tab:exp_statistics}
\end{table}

\begin{table*}[ht]
\setlength\tabcolsep{15pt}
\centering
\caption{Descriptions of the eight datasets.}
\begin{tabular}{lcccc}\hline
& Set A & Set B & Set C & Set D \\\hline \hline 
Facilities & 3 & 3 & 3 & 3 \\
Grids & 1000 $\times$ 1000 & 2000 $\times$ 2000 & 3000 $\times$ 3000 & 4000 $\times$ 4000 \\\hline\hline
& Set E & Set F & Set G & Set H\\\hline \hline 
Facilities & 3 & 4 & 5 & 6 \\
Grids & 5000 $\times$ 5000 & 3000 $\times$ 3000 & 3000 $\times$ 3000 & 3000 $\times$ 3000 \\\hline
\end{tabular}
\label{tab:data}
\end{table*}

\begin{table*}[ht]
\setlength\tabcolsep{18pt}
\centering
\caption{Descriptions of the tasks of the six experiments.}
\begin{tabular}{ll}\hline
Notation & Content\\\hline \hline 
GD Map & Map processing for distance calculation \\
GD Reduce & Reduce processing for distance calculation \\
MDT Map & Map processing for distance tuple creation \\
MDT Reduce & Reduce processing for distance tuple creation \\
GM Map & Map processing for central value calculation \\
GM Reduce & Reduce processing for central value calculation \\
LS Map & Map processing for local skyline calculation \\
LS Reduce & Reduce processing for local skyline calculation \\
GS Map & Map processing for global skyline calculation \\
GS Reduce & Reduce processing for global skyline calculation \\
MF & Creation of filters for the proposed algorithm \\
F & Implementing the filtering of the proposed algorithm \\
TS & Data sampling \\
MSQT & Creation of sky quadtree \\
MVM & Creation of virtual maximum point \\
MSF & Creation of Sky filter \\\hline
\end{tabular}
\label{tab:tasks}
\end{table*}

\subsection{Evaluation Datasets}
\label{sec:dataset}
Table \ref{tab:data} provides an overview of the properties of the eight datasets employed in our experiments. The datasets were carefully selected to evaluate the scalability of the proposed algorithm and assess the impact of the number of facilities. Five datasets, labeled Sets A to E, were used to analyze the scalability of the algorithm as the number of grids increased. These datasets incorporate three different types of facilities and services to assess the algorithm's performance under varying grid sizes. Furthermore, Set C, along with the other three datasets (Set F, Set G, and Set H), was deliberately designed with a consistent size of $3000\times 3000$ grids to evaluate the influence of the number of facilities. The facilities within each dataset were randomly distributed at a rate of one facility per row. In the case where the identical grid is occupied by multiple facilities, a tuple that is near the origin is introduced. This arrangement ensured that the proposed algorithm's filtering process would potentially eliminate a significant number of tuples, leading to improved results. To prevent such scenarios and maintain a fair evaluation, the datasets were created with overlapping ratios taken into account. Furthermore, considering that in real-world scenarios involving the computation of area skylines, usually the amount of facilities and the amount of grids tends to decrease as the grid size increases, we deliberately structured the datasets to capture this characteristic. Given that an increase in the number of grids might result in a decrease in the number of facilities, this experiment allows for a comprehensive evaluation of the algorithm's performance across various scenarios. By employing a diverse range of datasets with varying grid sizes, facility types, and distribution patterns, we aim to thoroughly assess the scalability, effectiveness, and robustness of the proposed algorithm in computing the area skyline.

\subsection{Baselines}
\label{sec:baselines}
The following six baselines are used in the experiments.
\begin{itemize}
\item E-MR-BNL: Existing algorithm + MR-BNL
\item E-MR-SFS: Existing algorithm + MR-SFS
\item E-SKY-MR: Existing algorithm + SKY-MR
\item P-MR-BNL: Proposed algorithm + MR-BNL
\item P-MR-SFS: Proposed algorithm + MR-SFS
\item P-SKY-MR: Proposed algorithm + SKY-MR
\end{itemize}

The area skyline computation process encompasses three key components: calculation of distance, creation of distance tuples, and the computation of skyline objects. In prior work, an established algorithm was utilized for the calculation of distance. As for the creation of distance tuples, various existing or proposed algorithms were employed. For the skyline calculation stage, the options included MR-BNL, MR-SFS, or SKY-MR algorithms.

\begin{figure*}[htbp]
\centering
\subfigure[Execution time with an increase in the number of grids.]{
\includegraphics[width=0.49\textwidth]{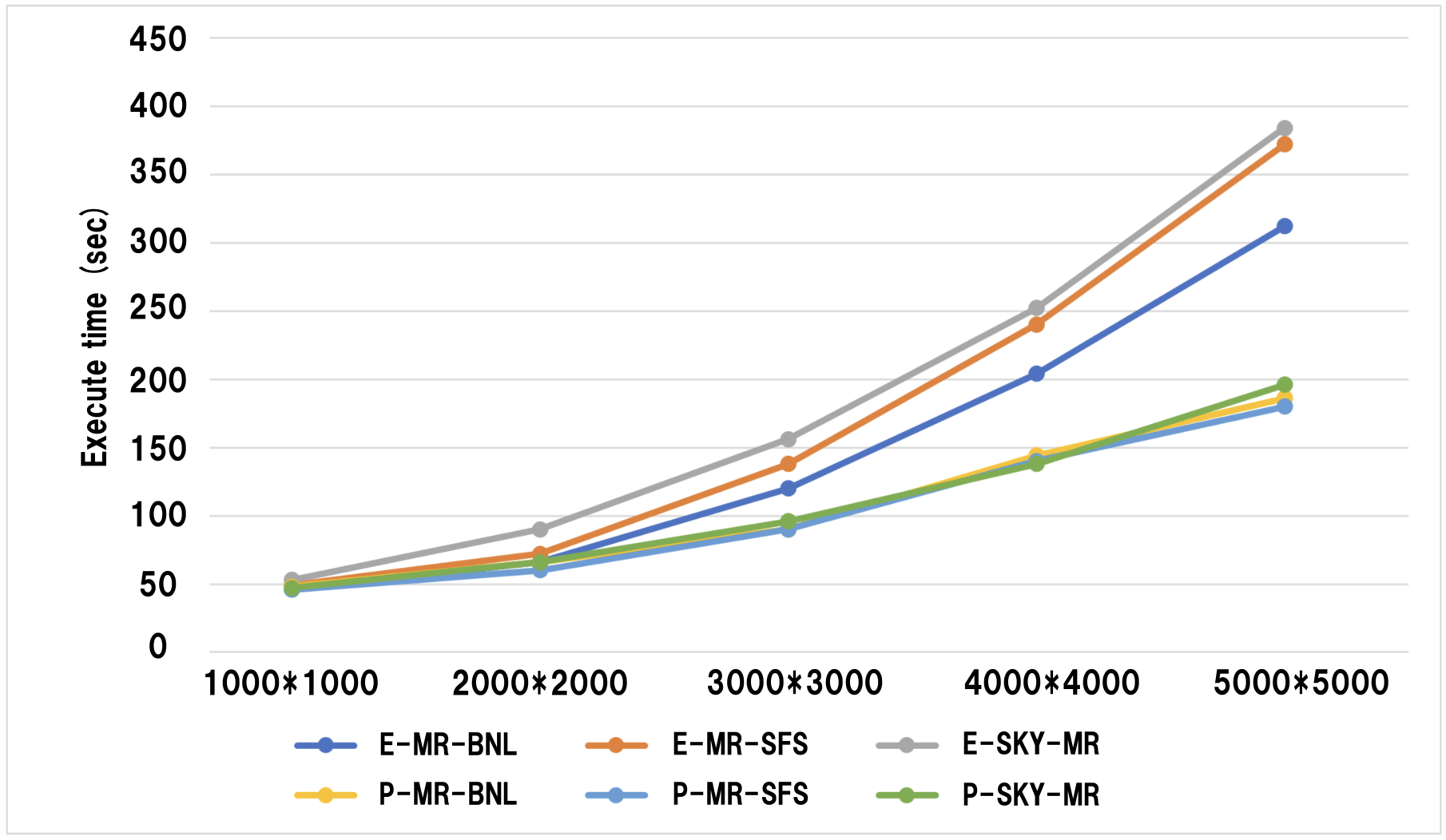}
\label{fig:exp_grids}
}\hspace{-3mm}
\subfigure[Execution time with an increase in the number of facilities.]{
\includegraphics[width=0.49\textwidth]{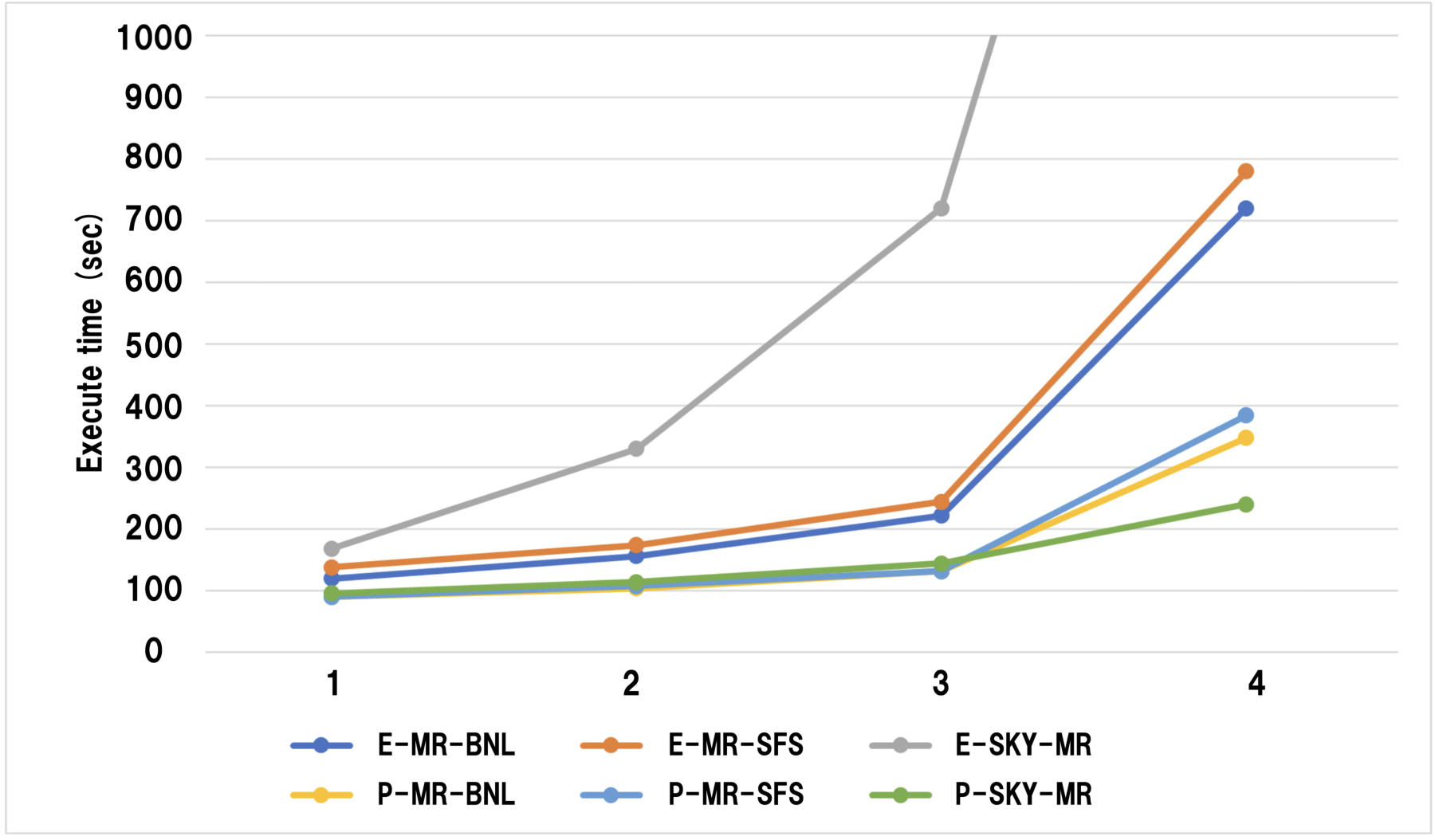}
\label{fig:exp_facilities}
}
\caption{Relationship between the number of grids and facilities and the execution time.}
\end{figure*}

\subsection{Evaluation Measure}
\label{sec:metrics}
For each of the eight types of experiments, the execution is carried out, and the average execution time is used as the evaluation metric.

\subsection{Tasks}
\label{sec:tasks}
Table \ref{tab:tasks} provides a description of the tasks performed in the six experiments.

\subsection{Effect of Grids}
\label{sec:grids}
Figure \ref{fig:exp_grids} depicts the correlation between the quantity of grids and the execution time. There is minimal difference in execution time between the existing algorithms and our presented algorithm when the grid dimensions are 1000 $\times$ 1000 and 2000 $\times$ 2000. However, with an increase in data size to 3000 $\times$ 3000 grids or larger, our algorithms showcase notable reductions in execution times. Specifically, for 3000 $\times$ 3000 data, the proposed algorithms achieve a reduction in execution time ranging from 62\% to 75\% compared to the existing methods. For 4000 $\times$ 4000 grids, the time is curtailed to 55\% to 71\% of the execution time for the existing algorithms. As the amount of grids reaches 5000 $\times$ 5000, the execution time further diminishes to 48\% to 60\% of the execution time of the existing algorithms. In summary, while the execution time generally increases with larger grid sizes, the proposed algorithms consistently manifest lower time consumption in comparison to the baseline algorithms.

\begin{figure*}[htbp]
\centering
\subfigure[E-MR-BNL]{
\includegraphics[width=0.49\textwidth]{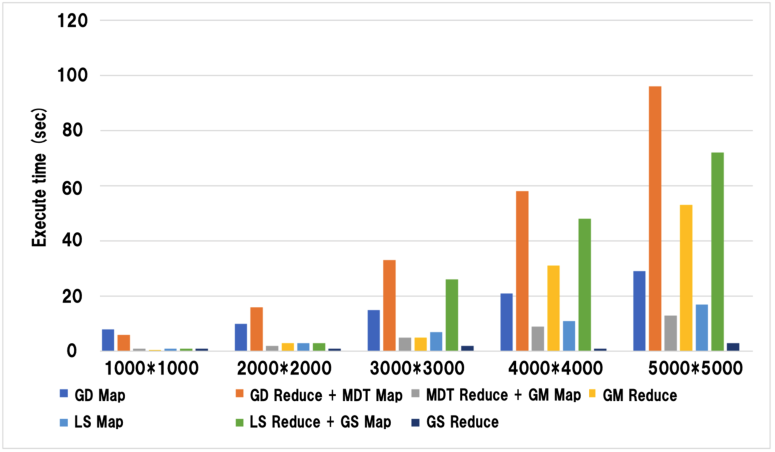}
}\hspace{-3mm}
\subfigure[P-MR-BNL]{
\includegraphics[width=0.49\textwidth]{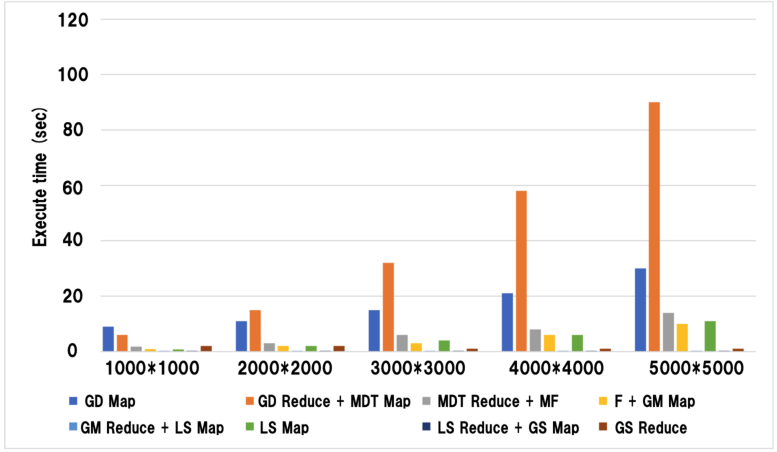}
}\hspace{-3mm}
\subfigure[E-MR-SFS]{
\includegraphics[width=0.49\textwidth]{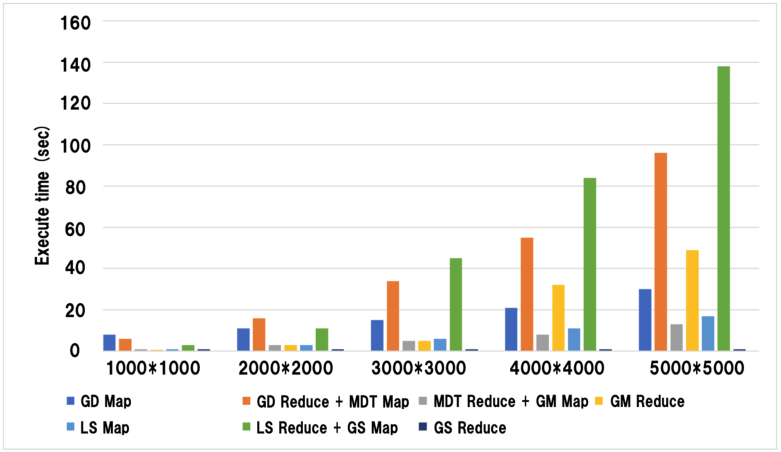}
}\hspace{-3mm}
\subfigure[P-MR-SFS]{
\includegraphics[width=0.49\textwidth]{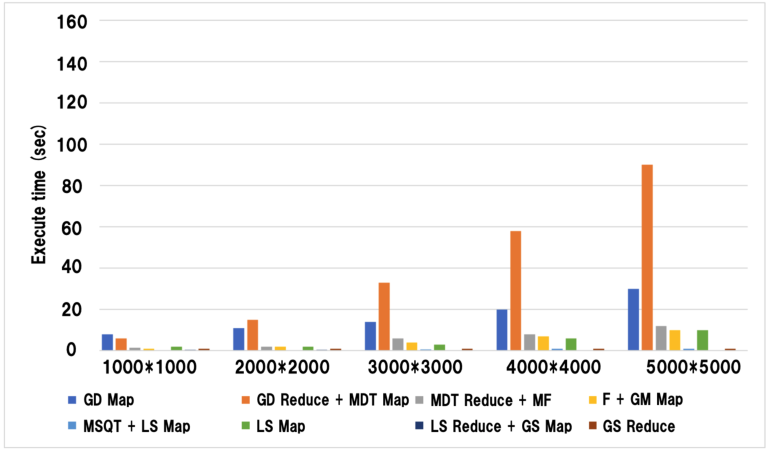}
}\hspace{-3mm}
\subfigure[E-SKY-MR]{
\includegraphics[width=0.49\textwidth]{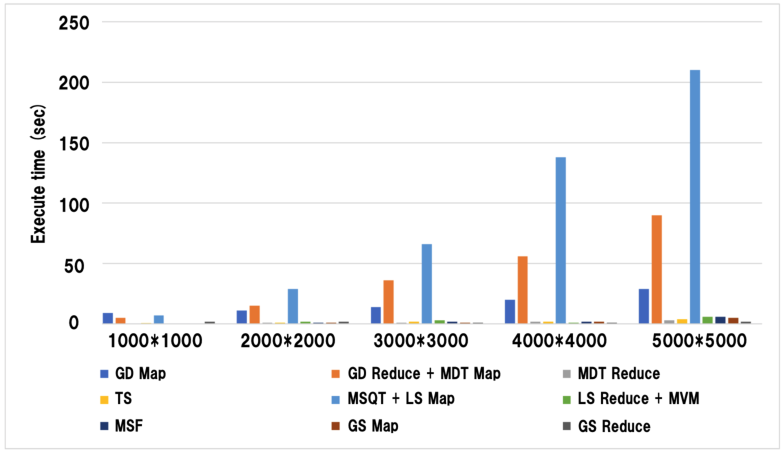}
}\hspace{-3mm}
\subfigure[P-SKY-MR]{
\includegraphics[width=0.49\textwidth]{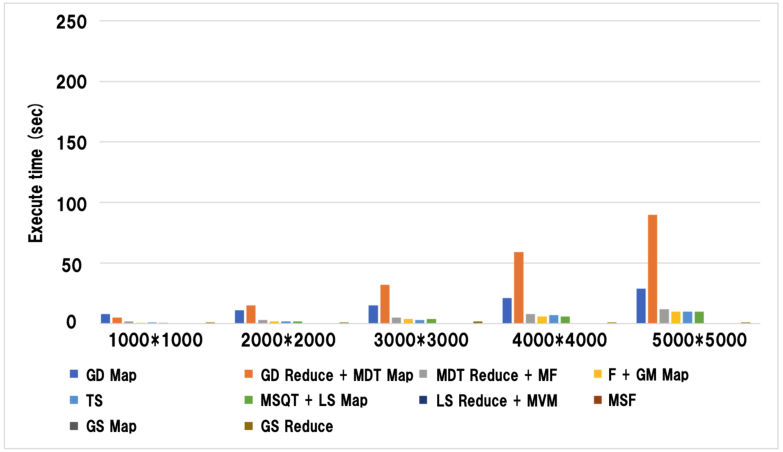}
}
\caption{Relationship between the number of grids per stage and execution time.}
\label{fig:ex_all}
\end{figure*}

\begin{figure*}[htbp]
\centering
\subfigure[E-MR-BNL]{
\includegraphics[width=0.49\textwidth]{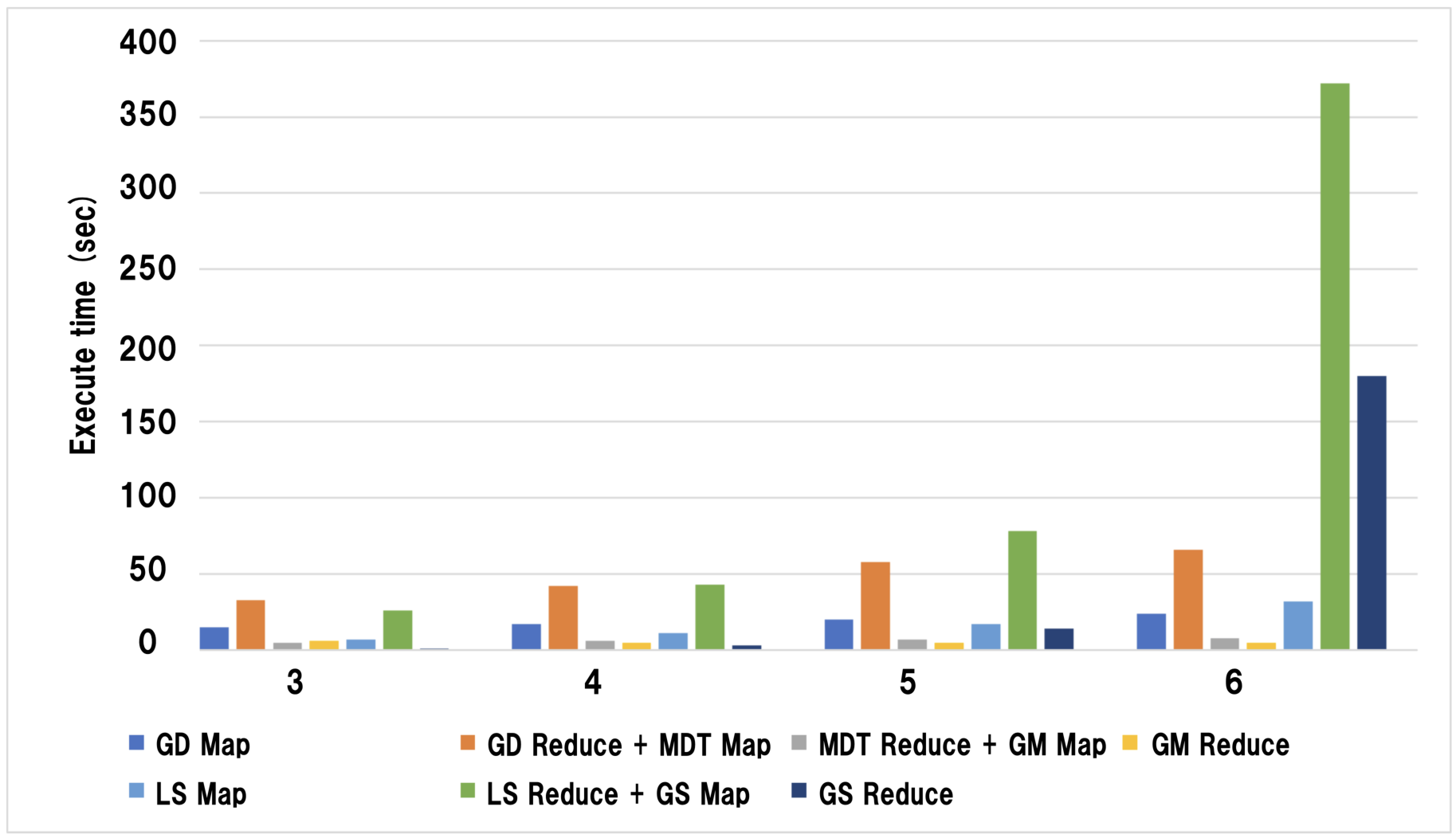}
}\hspace{-3mm}
\subfigure[P-MR-BNL]{
\includegraphics[width=0.49\textwidth]{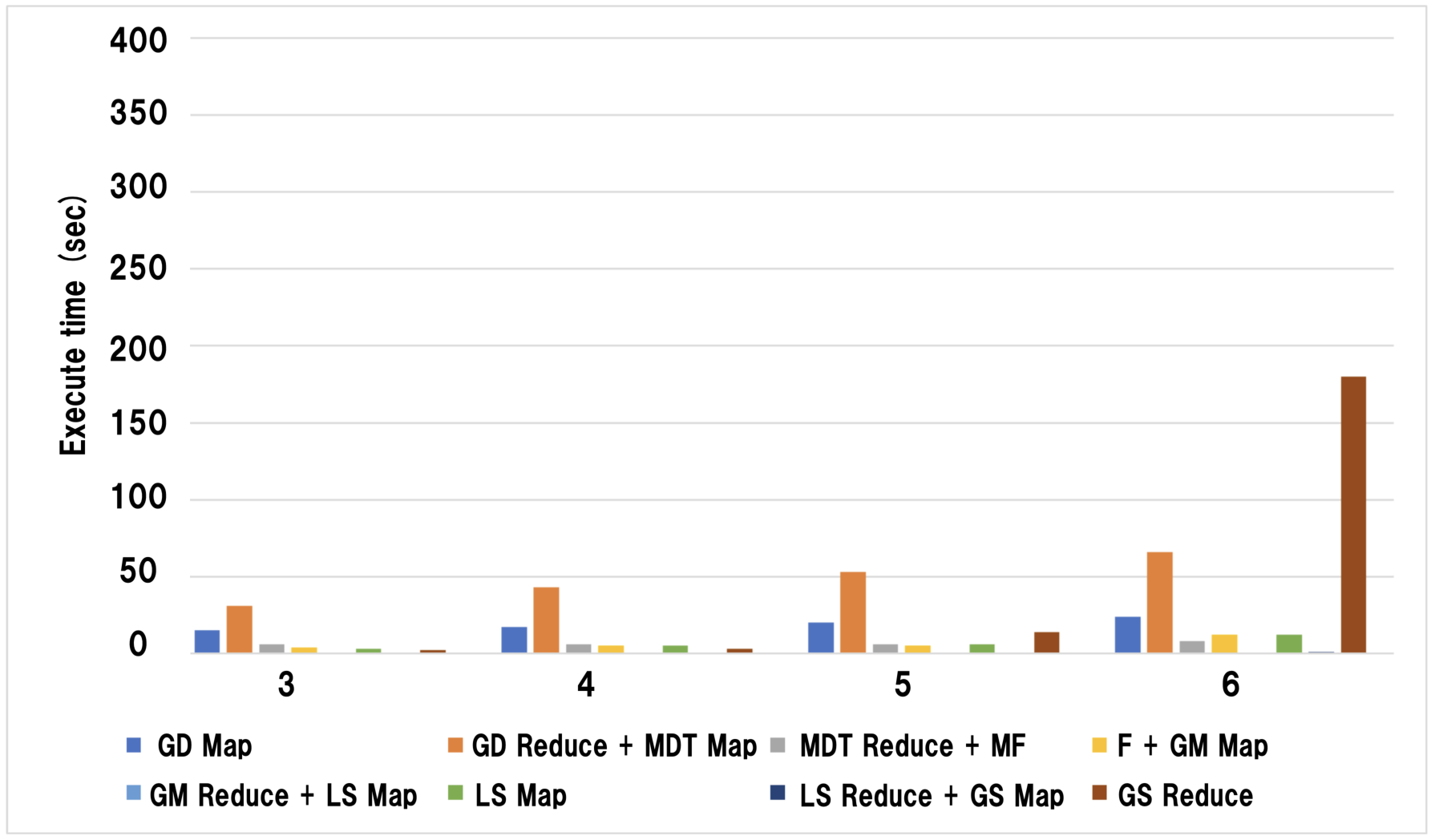}
}
\subfigure[E-MR-SFS]{
\includegraphics[width=0.49\textwidth]{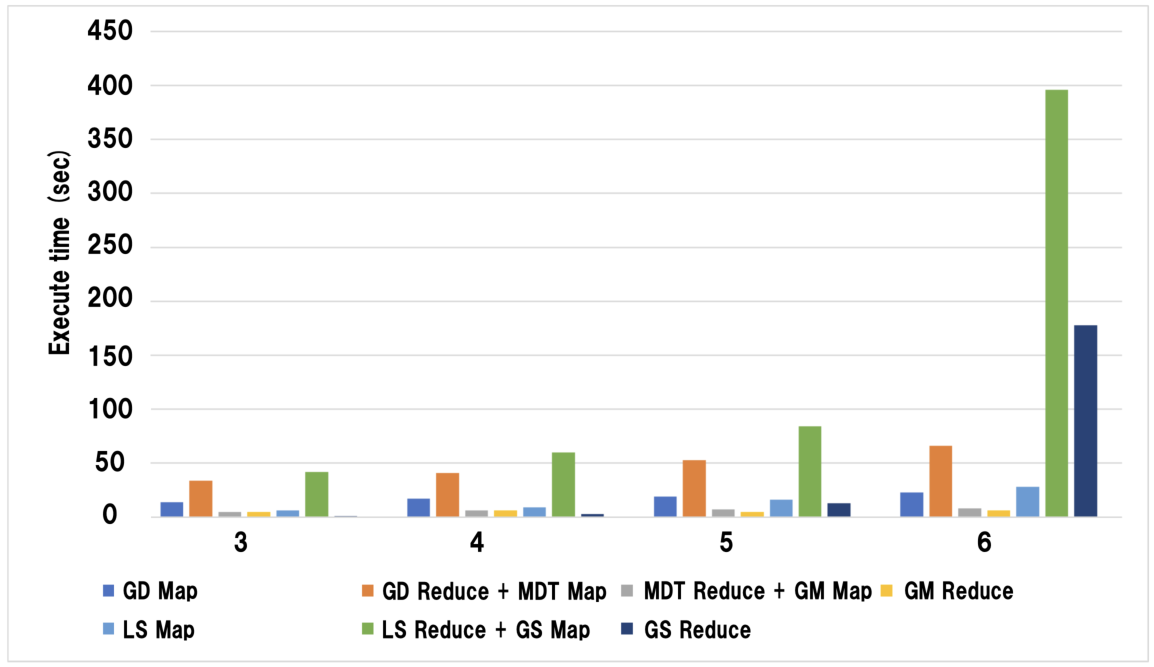}
}\hspace{-3mm}
\subfigure[P-MR-SFS]{
\includegraphics[width=0.49\textwidth]{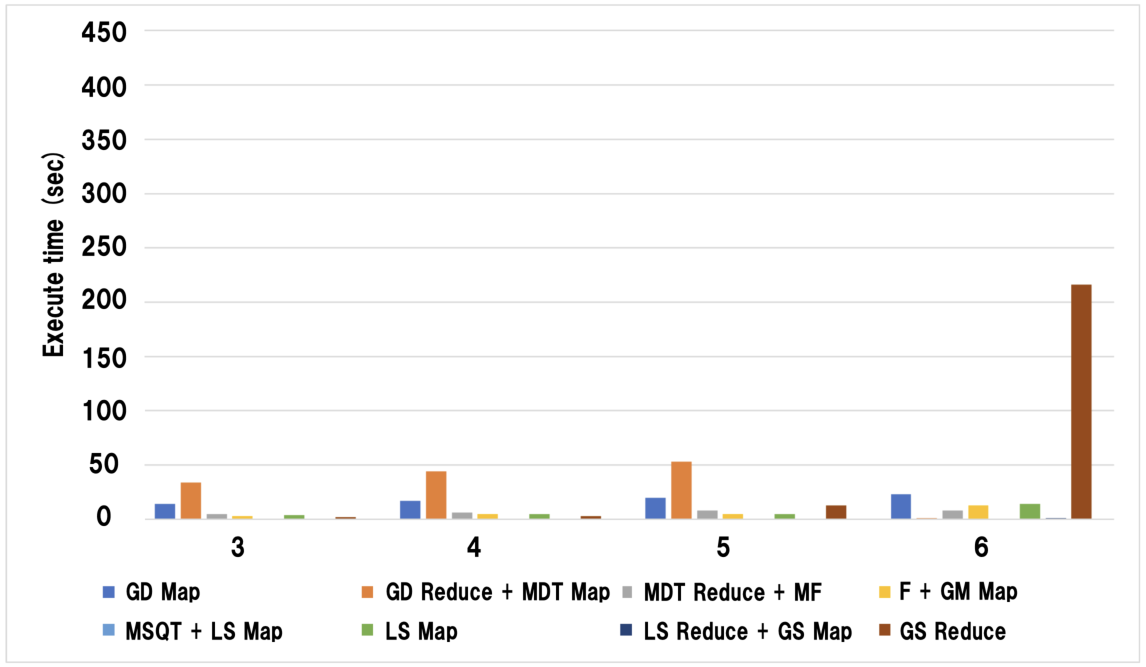}
}
\subfigure[E-SKY-MR]{
\includegraphics[width=0.49\textwidth]{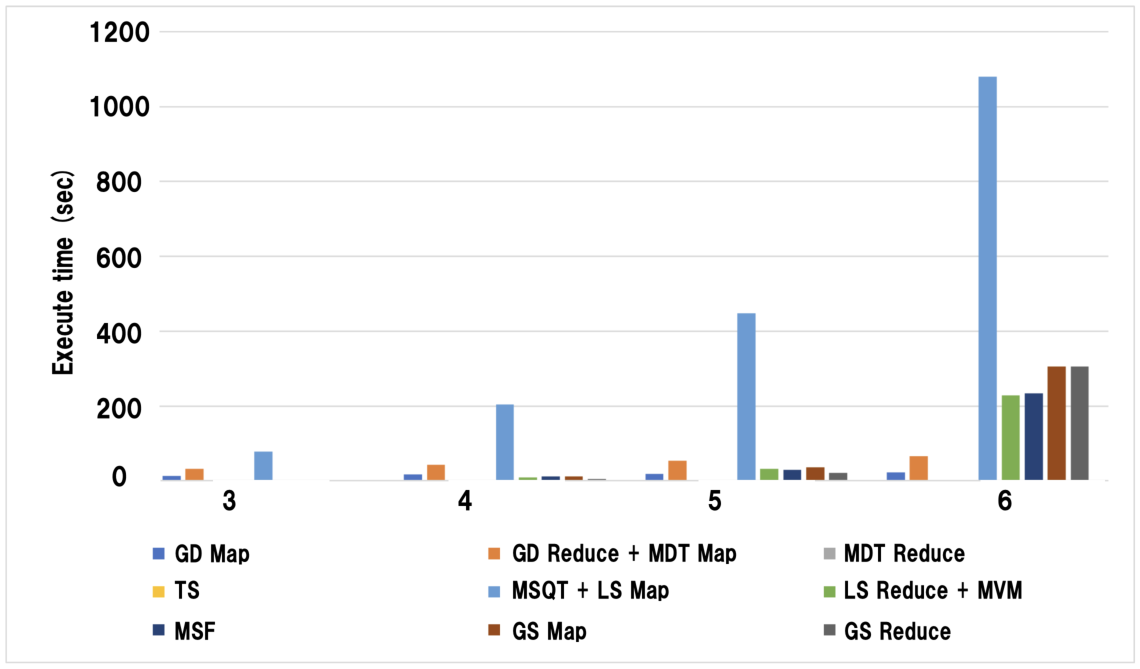}
}\hspace{-3mm}
\subfigure[P-SKY-MR]{
\includegraphics[width=0.49\textwidth]{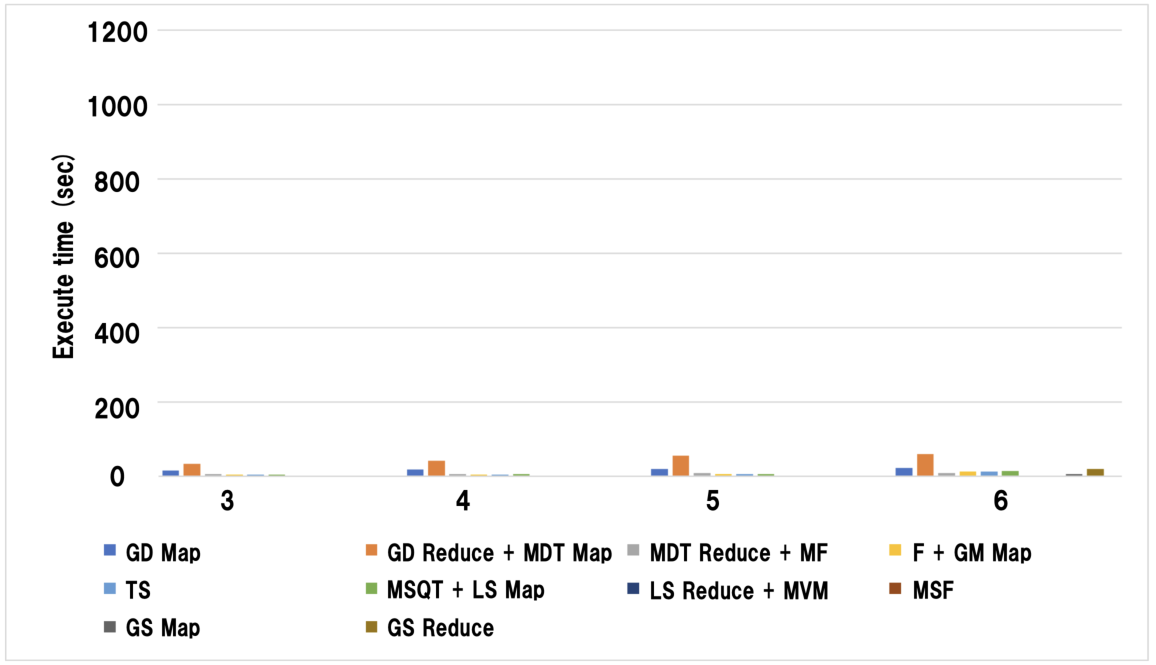}
}
\caption{Relationship between the number of facilities per stage and execution time.}
\label{fig:ex_all2}
\end{figure*}

Figure \ref{fig:ex_all} illustrates the correlation between the amount of grids and the time per stage. The proposed Apache Spark-based algorithm demonstrates a significant reduction in execution time at the skyline computation stage. In the case of MR-BNL and MR-SFS, the time to execute Reduce process in local skyline computation shows an increase with the growing number of grids. However, the proposed algorithm achieves a remarkable reduction in execution time, ranging from 0.1\% to 20\% compared to the existing algorithms. In SKY-MR, the execution time for sky quadtree creation exhibits an increase. However, due to the filtering of data size before skyline calculation in our proposed algorithm, the burden of tree creation is alleviated. As a result, the execution time of the proposed algorithm is reduced by 0.1\% to 20\%. Overall, the proposed algorithm consistently outperforms the existing algorithms, delivering significant time savings at the skyline computation stage.

\subsection{Effect of Facilities}
\label{sec:facilities}
Figure \ref{fig:exp_facilities} illustrates the relationship between the number of facilities and the overall execution time. Notably, the proposed Apache Spark-based algorithm exhibits significant reductions in execution time across a different number of facilities. When the number of facilities is 3, the execution time is reduced by an impressive 59\% to 75\% compared to the baseline algorithms. Furthermore, with four and five facilities, the execution time experiences substantial reductions of 35\% to 67\% and 20\% to 60\%, respectively. These findings indicate that as the number of facilities increases, the execution time tends to increase as well. However, our proposed algorithm consistently outperforms the baseline algorithms, showcasing lower time consumption and improved efficiency. The observed execution time reductions highlight the effectiveness of our approach in handling larger datasets with varying numbers of facilities. By leveraging optimized algorithms and techniques, our proposed solution offers substantial time savings and enhances the overall performance of area skyline computation.

Figure \ref{fig:ex_all2} provides insights into the relationship between the increasing number of facilities and the execution time of each stage in the area skyline computation process. Our proposed algorithm demonstrates notable reductions in the execution time, particularly in the skyline computation stage. In the MR-BNL and MR-SFS algorithms, the execution time of the reduced process of the local skyline calculation shows an upward trend as the number of facilities increases. However, our proposed method significantly mitigates this increase, achieving execution time reductions of approximately 0.2\% to 0.7\% compared to the existing algorithms. Similarly, in the SKY-MR algorithm, the execution time for the sky quadtree creation stage exhibits an increase with the growing number of facilities. Nonetheless, our proposed algorithm effectively curtails this escalation, resulting in execution time reductions of approximately 1\% to 5\% in comparison to the baseline algorithms. These findings highlight the efficacy of our proposed approach in improving the efficiency of skyline computation, even when faced with an increasing number of facilities. By leveraging optimized algorithms and techniques, our method ensures that the execution time remains manageable and allows for faster and more effective analysis of the area skyline.

\section{Conclusion}
\label{sec:conclusion}
The problem of area skyline computation entails identifying optimal locations for a set of query points based on multiple criteria. However, the computational complexity associated with this problem is substantial, and existing solutions lack efficiency and adaptability for distributed area skyline computation. To address this challenge, the study introduced a novel Apache Spark-based distributed algorithm aimed at minimizing the computation time of area skyline. The proposed algorithm incorporated three crucial techniques: filter creation from local partial skyline points during distance table creation, filter creation at the driver, and filtering in each executor. Implementation of these techniques effectively reduced the computational load associated with area skyline calculation, resulting in a notable decrease in overall execution time. It is noteworthy that the execution time in the proposed algorithm is directly impacted by two factors: the number of grids and the number of facilities.

It is important to note that this study focuses solely on evaluating the proposed algorithm using eight synthetic datasets. While these datasets serve as a starting point for analysis, they may not fully capture the complexities and nuances of real-world scenarios. Therefore, future research endeavors will involve applying the proposed algorithm to real-world datasets for case studies in various domains such as location-based services, spatial decision-making, and environmental monitoring. By examining the algorithm's performance in practical settings, a more comprehensive understanding of its effectiveness and applicability can be gained, leading to further improvements and optimizations.

\section*{Acknowledgement}
\label{sec:ack}
The preliminary version of this paper was published in Proceedings of the International Conference on Knowledge Science, Engineering and Management (KSEM 2023) \citep{li2023enhanced}. In this extended version, we have provided more details on the model’s architecture and implementation, and reported new experimental results on several benchmark datasets. 
\bibliographystyle{elsarticle-num}
\bibliography{refs}


\end{document}